\documentclass[letterpaper,12pt]{JHEP3}
\usepackage{amsmath,amssymb}
\usepackage{latexsym}
\usepackage{epsfig}
\usepackage{cite}
\usepackage[english]{babel}

\newcommand{\eq}{\begin{equation}}
\newcommand{\feq}{\end{equation}}
\newcommand{\eqn}{\begin{eqnarray}}
\newcommand{\feqn}{\end{eqnarray}}
\newcommand{\arr}{\begin{eqnarray*}}
\newcommand{\farr}{\end{eqnarray*}}

\font\mybb=msbm10 at 12pt
\def\bb#1{\hbox{\mybb#1}}
\def\bZ {\bb{Z}}

\def\bR {\bb{R}}

\def\bC {\bb{C}}

\title{The CFT dual of AdS gravity with torsion}
\author{Dietmar Klemm and Giovanni Tagliabue \\
Dipartimento di Fisica dell'Universit\`a di Milano \\
Via Celoria 16, I-20133 Milano and \\
INFN, Sezione di Milano, Via Celoria 16, I-20133 Milano. \\
E-mail: \email{dietmar.klemm@mi.infn.it,
               giovanni.tagliabue@mi.infn.it}}
\preprint{IFUM-894-FT}

\abstract{We consider the Mielke-Baekler model of three-dimensional AdS gravity
with torsion, which has gravitational and translational Chern-Simons terms in
addition to the usual Einstein-Hilbert action with cosmological constant. It is
shown that the topological nature of the model leads to a finite Fefferman-Graham
expansion. We derive the holographic stress tensor and the associated Ward
identities and show that, due to the asymmetry of the left- and right-moving
central charges, a Lorentz anomaly appears in the dual conformal field theory.
Both the consistent and the covariant Weyl and Lorentz anomaly are determined,
and the Wess-Zumino consistency conditions for the former are verified.
Moreover we consider the most general solution with flat boundary geometry,
which describes left-and right-moving gravitational waves on AdS$_3$ with torsion,
and shew that in this case the holographic energy-momentum tensor is given by the
wave profiles. The anomalous transformation laws of the wave profiles under
diffeomorphisms preserving the asymptotic form of the bulk solution yield the central
charges of the dual CFT and confirm the results that appeared earlier on in the
literature. We finally comment on some points concerning the microstate counting
for the Riemann-Cartan black hole.}


\begin{document}

\section{Introduction}
\label{intro}

According to the AdS/CFT correspondence (cf.~\cite{Aharony:1999ti} for a review),
any theory of gravity on a $d+1$-dimensional asymptotically anti-de~Sitter space
is dual to a conformal field theory living on the $d$-dimensional boundary of AdS.
This allows to compute CFT correlation functions of operators $\cal O$ by considering
fields $\phi$ propagating in the $d+1$-dimensional bulk spacetime. The boundary
value $\phi_0$ of $\phi$ represents a source for the associated operator $\cal O$.
By turning on various bulk fields one can deform the corresponding CFT, and break
symmetry explicitely or spontaneously, depending on the boundary condition on $\phi$.\\
A generalization that has not been investigated very much up to now is to admit torsion in
the gravity theory\footnote{The holographic currents associated to five-dimensional
Chern-Simons gravity with nonvanishing torsion were studied in \cite{Banados:2006fe}.},
and to address this point is the purpose of the present paper.
We will study the effects of torsion in a simple setting, represented by a
topological model of three-dimensional gravity, whose equations of motion imply
both constant curvature and constant torsion \cite{Mielke:1991nn}. What makes this
model particularly appealing is the fact that, similar to ordinary three-dimensional
general relativity with negative cosmological constant, it can be written as a sum of
two SL$(2,\bR)$ Chern-Simons theories, but with unequal coupling
constants \cite{Blagojevic:2003vn,Cacciatori:2005wz}. We derive the
central charges of the dual CFT, the holographic energy-momentum tensor and the
associated (anomalous) Ward identities. In particular, there is a Lorentz anomaly,
which comes from the presence of a gravitational Chern-Simons term in the bulk action,
invariant under local Lorentz transformations only up to a boundary term.
The holographic description of diffeomorphism and Lorentz anomalies by gravitational
Chern-Simons terms was explored in \cite{Kraus:2005zm}. We find that bulk torsion
modifies the trace anomaly, but the Lorentz anomaly is given by the prefactor of
the gravitational Chern-Simons term alone.\\
Our paper is organized as follows: In section \ref{grav-torsion} we briefly review
the Mielke-Baekler model of three-dimensional gravity with torsion, and its formulation
as a Chern-Simons theory. In the following section we work out the Fefferman-Graham
expansion for the dreibein and the spin connection and show that it is finite.
In section \ref{Hst}, the holographic stress tensor and the associated
anomalous Ward identities are obtained. We determine both the consistent
and the covariant anomalies, as well as the Bardeen-Zumino polynomial
relating them. It is furthermore shown that no diffeomorphism
(Einstein) anomaly appears. We then consider the most general bulk solution with flat
boundary, which represents left-and right-moving gravitational waves on AdS$_3$ with
torsion. In this case the CFT energy-momentum tensor reduces to the wave profiles,
and transforms anomalously under diffeomorphisms preserving the asymptotic form of
the solution. From the transformation laws one can read off the central charges,
and confirm the results of \cite{Blagojevic:2004hj}.
Finally, in section \ref{entr-RC} we discuss some points related to the microstate
counting for the Riemann-Cartan black hole.
In the appendix, we check that our anomalies satisfy the Wess-Zumino consistency
conditions.

\section{Three-dimensional gravity with torsion}
\label{grav-torsion}

A simple three-dimensional
model that yields nonvanishing torsion was proposed by Mielke and Baekler
(MB) \cite{Mielke:1991nn} and further analyzed by Baekler, Mielke and
Hehl \cite{Baekler:1992ab}. The action reads \cite{Mielke:1991nn}\footnote{Our
conventions are as follows: $A,B,\ldots$ are 3d Lorentz indices, while
$\mu,\nu,\ldots$ are 3d spacetime indices. Two-dimensional Lorentz and world
indices on the boundary of AdS$_3$ are denoted by $a,b,\ldots$ and $i,j,\ldots$
respectively. The signature is mostly plus, and hatted fields are objects in
three dimensions.}
\begin{equation}
I = a I_1 + \Lambda I_2 + \alpha_3 I_3 + \alpha_4 I_4\,, \label{MBaction}
\end{equation}
where $a$, $\Lambda$, $\alpha_3$ and $\alpha_4$ are constants,
\begin{eqnarray}
I_1 &=& 2\int {\hat e}_A \wedge {\hat R}^A\,, \nonumber \\
I_2 &=& -\frac 13 \int \epsilon_{ABC} {\hat e}^A \wedge {\hat e}^B \wedge {\hat e}^C\,,
        \nonumber \\
I_3 &=& \int {\hat \omega}_A \wedge d{\hat \omega}^A + \frac 13 \epsilon_{ABC}
        {\hat \omega}^A \wedge {\hat \omega}^B \wedge {\hat \omega}^C\,, \nonumber \\
I_4 &=& \int {\hat e}_A \wedge {\hat T}^A\,, \nonumber
\end{eqnarray}
and
\begin{eqnarray}
{\hat R}^A &=& d{\hat \omega}^A + \frac 12 {\epsilon^A}_{BC}\,{\hat \omega}^B \wedge
{\hat \omega}^C\,, \nonumber \\
{\hat T}^A &=& d{\hat e}^A + {\epsilon^A}_{BC}\,{\hat \omega}^B \wedge {\hat e}^C\,,
\label{1stCartan}
\end{eqnarray}
denote the curvature and torsion two-forms respectively.
${\hat \omega}^A$ is defined by
${\hat \omega}^A = \frac 12 \epsilon^{ABC}{\hat \omega}_{BC}$ with $\epsilon_{012} = 1$.
$I_1$ yields the Einstein-Hilbert
action, $I_2$ a cosmological constant, $I_3$ is a Chern-Simons term for the spin
connection\footnote{Some aspects of three-dimensional gravity with gravitational
Chern-Simons term were studied in \cite{Park:2006gt}.}, and $I_4$ represents
a translational Chern-Simons term.
Note that, in order to obtain the topologically massive gravity of Deser, Jackiw and
Templeton (DJT) \cite{Deser:1981wh} from (\ref{MBaction}), one has to add a Lagrange
multiplier term that ensures vanishing torsion.
The field equations following from (\ref{MBaction}) take the form
\begin{eqnarray}
2a{\hat R}^A - \Lambda {\epsilon^A}_{BC}\,{\hat e}^B \wedge {\hat e}^C + 2\alpha_4
{\hat T}^A &=& 0\,, \nonumber \\
2a{\hat T}^A + 2\alpha_3 {\hat R}^A + \alpha_4 {\epsilon^A}_{BC}\,{\hat e}^B \wedge
{\hat e}^C &=& 0\,. \nonumber
\end{eqnarray}
In what follows, we assume $\alpha_3\alpha_4 - a^2 \neq 0$\footnote{For
$\alpha_3\alpha_4 - a^2 = 0$ the theory becomes singular \cite{Baekler:1992ab}.}.
Then the equations of motion can be rewritten as
\begin{equation}
2{\hat T}^A = A{\epsilon^A}_{BC}\,{\hat e}^B \wedge {\hat e}^C\,, \qquad
2{\hat R}^A = B{\epsilon^A}_{BC}\,{\hat e}^B \wedge {\hat e}^C\,, \label{eomMB}
\end{equation}
where
\begin{displaymath}
A = \frac{\alpha_3\Lambda + \alpha_4 a}{\alpha_3\alpha_4 - a^2}\,, \qquad
B = -\frac{a\Lambda+\alpha_4^2}{\alpha_3\alpha_4 - a^2}\,.
\end{displaymath}
Thus, the field configurations are characterized by constant curvature and
constant torsion. From \eqref{1stCartan} one gets
\begin{equation}
{\hat \omega}^A = {\hat\omega}^{(0)A} - {\hat K}^A\,, \label{decomp_omega}
\end{equation}
where ${\hat\omega}^{(0)A}$ denotes the
Christoffel connection and ${\hat K}^A$ is the contorsion one-form given by
${\hat K}^A_{\mu} = \frac 12 {\epsilon^A}_{BC}\,{\hat e}^{B\beta}
{\hat e}^{C\gamma}{\hat K}_{\beta\gamma\mu}$,
with the contorsion tensor
\begin{displaymath}
{\hat K}_{\beta\gamma\mu} = \frac 12\left({\hat T}_{\beta\gamma\mu} -
{\hat T}_{\gamma\beta\mu} - {\hat T}_{\mu\beta\gamma}\right)\,,
\end{displaymath}
and ${\hat T}_{\beta\gamma\mu} = {\hat e}_{A\beta}{\hat T}^A_{\;\;\gamma\mu}$.
\eqref{decomp_omega}
allows to express the curvature ${\hat R}^A$ of a Riemann-Cartan
spacetime in terms of its Riemannian part ${\hat R}^{(0)A}$ and ${\hat K}^A$,
\begin{equation}
{\hat R}^A = {\hat R}^{(0)A} - d{\hat K}^A - {\epsilon^A}_{BC}\,{\hat \omega}^B
\wedge {\hat K}^C - \frac 12{\epsilon^A}_{BC}\,{\hat K}^B \wedge {\hat K}^C\,. \label{RK}
\end{equation}
Using the equations of motion (\ref{eomMB}) in (\ref{RK}),
one gets for the Riemannian part
\begin{equation}
2{\hat R}^{(0)A} = \Lambda_{\mathrm{eff}}{\epsilon^A}_{BC}\,{\hat e}^B\wedge {\hat e}^C\,,
                   \label{Riem-curv}
\end{equation}
with the effective cosmological constant
\begin{displaymath}
\Lambda_{\mathrm{eff}} = B - \frac{A^2}4\,.
\end{displaymath}
This means that locally the metric is given by the (anti-)de~Sitter or Minkowski
solution, depending on whether $\Lambda_{\mathrm{eff}}$ is negative, positive or
zero. It is interesting to note that $\Lambda_{\mathrm{eff}}$ can be nonvanishing
even if the bare cosmological constant $\Lambda$ is zero \cite{Baekler:1992ab}.
In this simple model, dark energy (i.~e.~, $\Lambda_{\mathrm{eff}}$) would then be
generated by the translational Chern-Simons term $I_4$.

In \cite{Blagojevic:2003vn} it was shown that for $\Lambda_{\mathrm{eff}} < 0$, the
Mielke-Baekler model (\ref{MBaction}) can be written as a sum of two SL$(2,\bR)$
Chern-Simons theories. This was then generalized in \cite{Cacciatori:2005wz}
to the case of arbitrary effective cosmological constant.
In what follows we shall be interested in the case $\Lambda_{\mathrm{eff}} < 0$,
so we briefly summarize the results of \cite{Blagojevic:2003vn}. For
$\Lambda_{\mathrm{eff}} < 0$ the geometry is locally AdS$_3$, which has the
isometry group SO$(2,2)$ $\cong$ SL$(2,\bR) \times$ SL$(2,\bR)$, so if the MB model
is equivalent to a Chern-Simons theory, one expects a gauge group SO$(2,2)$.
Indeed, if one defines the SL$(2,\bR)$ connections
\begin{displaymath}
A^A = {\hat \omega}^A + q\, {\hat e}^A\,, \qquad {\tilde A}^A = {\hat \omega}^A +
\tilde q\, {\hat e}^A\,,
\end{displaymath}
then the SL$(2,\bR) \times$ SL$(2,\bR)$ Chern-Simons action\footnote{In (\ref{CSMBAdS}),
$\langle\tau_A\,,\tau_B\rangle = 2\,\mathrm{Tr}\,(\tau_A\tau_B) = \eta_{AB}$, and the
SL$(2,\bR)$ generators $\tau_A$ satisfy $[\tau_A, \tau_B] = {\epsilon_{AB}}^C\tau_C$.}
\begin{equation}
I_{CS} = \frac t{8\pi} \int \langle A \wedge dA + \frac 23 A \wedge A \wedge A\rangle -
         \frac{\tilde t}{8\pi}\int \langle\tilde A \wedge d\tilde A + \frac 23 \tilde A
         \wedge \tilde A \wedge \tilde A\rangle
         \label{CSMBAdS}
\end{equation}
coincides (up to boundary terms) with $I$ in (\ref{MBaction}), if the parameters
$q, \tilde q$ and the coupling constants $t, \tilde t$ are given by
\begin{equation}
q = -\frac A2 + \sqrt{-\Lambda_{\mathrm{eff}}}\,, \qquad
\tilde q = -\frac A2 - \sqrt{-\Lambda_{\mathrm{eff}}}\,
\end{equation}
and
\begin{equation}
\frac t{2\pi} = 2\alpha_3 + \frac{2a + \alpha_3 A}{\sqrt{-\Lambda_{\mathrm{eff}}}}\,,
                \qquad\frac{\tilde t}{2\pi} = -2\alpha_3 + \frac{2a + \alpha_3 A}
                {\sqrt{-\Lambda_{\mathrm{eff}}}}\,. \label{t}
\end{equation}
We see that $q, \tilde q$, and thus the connections $A^A, {\tilde A}^A$ are real for
negative $\Lambda_{\mathrm{eff}}$. The coupling constants $t, \tilde t$ are also real,
but in general different from each other due to the presence of $I_3$.

\section{Finite Fefferman-Graham expansion}

Let us now determine the Fefferman-Graham (FG) expansion \cite{FG} for the
dreibein ${\hat e}^A$ and the spin connection ${\hat \omega}^A$, which will turn
out to be finite\footnote{The fact that three-dimensional Einstein spaces with
negative curvature have a finite FG expansion was first shown
in \cite{Skenderis:1999nb}.}. To this end, we proceed similar
to \cite{Banados:2002ey,Banados:2006fe}, using the CS formulation of the MB model.
First of all, one assumes that the manifold is diffeomorphic to $M_2\times\bR$
asymptotically and that it is parametrized by the local coordinates
$x^{\mu} = (x^i, \rho)$, with $\rho$ denoting the radial coordinate and $M_2$
being the spacetime on which the dual CFT resides. The corresponding Lorentz indices
are split as $A=(a,2)$. The field equations $F=\tilde F=0$
following from \eqref{CSMBAdS} imply
\begin{equation}
\partial_{\rho} A_i - \partial_i A_{\rho} + [A_{\rho}, A_i] = 0\,, \label{dAi}
\end{equation}
and an analogous equation for $\tilde A$. Note that the simplest gauge choice
$A_{\rho}={\tilde A}_{\rho}=0$ is not allowed, as this would lead to a degenerate
dreibein. A nondegenerate choice is to take $A_{\rho}$ and ${\tilde A}_{\rho}$
to be constant Lie algebra elements. The general solution of \eqref{dAi} is then
given by
\begin{equation}
A_i(\rho, x^j) = e^{-\rho A_{\rho}}\,A_i(0, x^j)\,e^{\rho A_{\rho}}\,.
\label{Ai}
\end{equation}
As in \cite{Banados:2002ey} we choose $A_{\rho}=\tau_2$,
${\tilde A}_{\rho}=-\tau_2$, so that \eqref{Ai} leads to
\begin{eqnarray}
A_i(\rho, x^j) &=& A_i^0(0,x)(\tau_0\cosh\rho-\tau_1\sinh\rho)+A_i^1(0,x)
                   (\tau_1\cosh\rho-\tau_0\sinh\rho)+A_i^2(0,x)\tau_2\,, \nonumber \\
{\tilde A}_i(\rho, x^j) &=& {\tilde A}_i^0(0,x)(\tau_0\cosh\rho+\tau_1\sinh\rho)
                   +{\tilde A}_i^1(0,x)(\tau_1\cosh\rho+\tau_0\sinh\rho)+
                   {\tilde A}_i^2(0,x)\tau_2\,. \nonumber
\end{eqnarray}
Next, we shall impose one extra condition on the vielbein, namely
${\hat e}^2_{\;\;i}=0$, or equivalently $A_i^2(0,x)={\tilde A}_i^2(0,x)$.
This breaks three-dimensional Lorentz symmetry down to a two-dimensional one,
and leaves a 2d tetrad as a gravitational source. Moreover, it ensures that
the boundary metric is torsion-free \cite{Banados:2002ey}.
One obtains then the finite FG expansion
\begin{eqnarray}
{\hat e}^a(\rho, x) &=& e^{\rho}e^a(x) + e^{-\rho}k^a(x)\,, \nonumber \\
{\hat e}^2(\rho, x) &=& \ell d\rho\,, \nonumber \\
{\hat\omega}^a(\rho, x) &=& e^{\rho}\left\{\frac A2 e^a(x) + \frac 1{\ell}
{\epsilon^a}_b\,e^b(x)\right\} + e^{-\rho}\left\{\frac A2 k^a(x) -
\frac 1{\ell}{\epsilon^a}_b\,k^b(x)\right\}\,, \nonumber \\
{\hat\omega}^2(\rho, x) &=& \omega(x) + \frac{A\ell}2 d\rho\,, \label{FG}
\end{eqnarray}
for the dreibein and the spin connection, with $\ell$ defined by
$\Lambda_{\mathrm{eff}}=-1/\ell^2$, $\epsilon_{01}=1$, $\omega_i(x)=A^2_i(0,x)$ and
\begin{eqnarray}
{e^a}_i &=& \frac\ell4\left(A^a_i(0,x)-{\tilde A}^a_i(0,x)\right)+\frac\ell4
{\epsilon^a}_b\left(A^b_i(0,x)+{\tilde A}^b_i(0,x)\right)\,, \nonumber \\
{k^a}_i &=& \frac\ell4\left(A^a_i(0,x)-{\tilde A}^a_i(0,x)\right)-\frac\ell4
{\epsilon^a}_b\left(A^b_i(0,x)+{\tilde A}^b_i(0,x)\right)\,. \nonumber
\end{eqnarray}
$e^a$ and $\omega^{ab}=-\epsilon^{ab}\omega$ represent the tetrad and the spin
connection on the CFT manifold $M_2$.
Finally, the FG expansion of the three-dimensional line element
${\hat e}^A_{\;\;\mu}{\hat e}_{A\nu}dx^{\mu}dx^{\nu}$ is given by
\begin{equation}
d{\hat s}^2 = \left[e^{2\rho}g_{ij}+2k_{(ij)}+e^{-2\rho}\eta_{ab}{k^a}_i{k^b}_j\right]
dx^i dx^j + \ell^2d\rho^2\,,
\end{equation}
where $g_{ij}=\eta_{ab}{e^a}_i{e^b}_j$ and $k_{ij}=e_{ai}{k^a}_j$.
Note that the equations of motion \eqref{eomMB} for ${\hat T}^a$ imply
\begin{equation}
dk^a-{\epsilon^a}_b\,\omega\wedge k^b=0\,, \label{dk}
\end{equation}
as well as
\begin{equation}
de^a-{\epsilon^a}_b\,\omega\wedge e^b=0\,, \label{tors=0}
\end{equation}
i.~e.~the boundary torsion indeed vanishes. \eqref{eomMB} for ${\hat T}^2$ gives
furthermore $k_{[ij]}=0$, whereas ${\hat R}^2$ yields
\begin{equation}
d\omega + \frac 2{\ell^2}\epsilon_{ab}\,e^a\wedge k^b=0\,, \label{curv}
\end{equation}
and the field equation for ${\hat R}^a$ is identically satisfied.

\section{Holographic stress tensor}
\label{Hst}

In order to find the holographic energy-momentum tensor, we vary the action
\eqref{MBaction} on-shell, to get
\begin{equation}
\delta I = \int_{M_2}\left[-2a\,{\hat e}^A\wedge\delta{\hat\omega}_A-\alpha_3\,
           {\hat\omega}^A\wedge\delta{\hat\omega}_A-\alpha_4\,{\hat e}^A\wedge
           \delta{\hat e}_A\right]\,.
\end{equation}
Next, we evaluate this variation on the asymptotic solution \eqref{FG}. One finds
that the only divergent term in the limit $\rho\to\infty$ is given by
\begin{displaymath}
\delta I_{\mathrm{div}} = -\frac{2a}{\ell}\int e^{2\rho}\epsilon_{ab}\,e^a\wedge\delta e^b\,.
\end{displaymath}
This can be removed by adding to the action a local counterterm
\begin{displaymath}
I_{\mathrm{ct}} = \frac a{\ell}\int\epsilon_{ab}\,{\hat e}^a\wedge {\hat e}^b\,,
\end{displaymath}
which is the usual counterterm needed to regularize AdS$_3$
gravity \cite{Balasubramanian:1999re}\footnote{Note that $a=1/16\pi G$.}.
Up to terms that cancel in the limit $\rho\to\infty$ one gets then
\begin{eqnarray}
\delta(I+I_{\mathrm{ct}}) &=& -\frac{2\alpha_3}{\ell^2}\int e_a\wedge\delta k^a
+\left(\frac{4a}{\ell}+\alpha_3\frac A{\ell}\right)\int\epsilon_{ab}\,e^a\wedge\delta k^b
\nonumber \\
&& -\frac{2\alpha_3}{\ell^2}\int k_a\wedge\delta e^a - \alpha_3\frac A{\ell}\int
\epsilon_{ab}\,k^a\wedge\delta e^b - \alpha_3\int\omega\wedge\delta\omega\,. \nonumber
\end{eqnarray}
The next step is to transform variations of $k^a$ into variations of $e^a$. Up to
finite boundary terms, that we are free to add, one has
\begin{displaymath}
e_a\wedge\delta k^a = k_a\wedge\delta e^a\,,
\end{displaymath}
and a similar expression for $\epsilon_{ab}\,e^a\wedge\delta k^b$. In this way, we
finally arrive at
\begin{equation}
\delta I_{\mathrm{tot}} = -\frac{4\alpha_3}{\ell^2}\int
k_a\wedge\delta e^a - \left(\frac{4a}{\ell}+2\alpha_3\frac A{\ell}\right)\int
\epsilon_{ab}\,k^a\wedge\delta e^b - \alpha_3\int\omega\wedge\delta\omega\,,
\label{deltaItot}
\end{equation}
where $I_{\mathrm{tot}}=I+I_{\mathrm{ct}}+I_{\mathrm{fin.bdry.}}$.
One can now define the holographic energy-momentum tensor by\footnote{In
\eqref{stress-tens}, $\epsilon^{ij}$ is defined by $\epsilon^{tx}=-1$, if $t,x$ are
local coordinates on $M_2$. The orientation is such that
$dx^i\wedge dx^j=-\epsilon^{ij}d^2x$, and the Hodge dual is defined by
$(^*\omega)^i=|e|^{-1}\epsilon^{ij}\omega_j$. $\nabla_j$ denotes the covariant
derivative on $M_2$.}
\begin{equation}
{T^i}_a = \frac{2\pi}{|e|}\frac{\delta I_{\mathrm{tot}}}{\delta {e^a}_i}=
\frac{2\pi\epsilon^{ij}}{|e|}\left[-\frac{4\alpha_3}{\ell^2}k_{aj} + \frac2{\ell}
\left[2a+\alpha_3 A\right]\epsilon_{ab}{k^b}_j + \alpha_3 e_{am}\nabla_j
(^*\omega)^m\right]\,. \label{stress-tens}
\end{equation}
As was said earlier, the boundary torsion is zero, and thus the spin connection
$\omega$ is determined completely by $e^a$. This means that $\delta\omega$ in
\eqref{deltaItot} has to be expressed in terms of $\delta e^a$, and contributes
to the stress tensor\footnote{If the tetrad and the spin connection were independent,
the last term in \eqref{deltaItot} would not contribute to the stress tensor, but
would give rise to a spin current $\sigma^i=|e|^{-1}\delta I_{\mathrm{tot}}/\delta
\omega_i$. In five dimensions, such a scenario was considered in \cite{Banados:2006fe}.}.
Note also that ${T^i}_a$ is the Hodge dual of the energy-momentum one-form $\tau_a$,
${T^i}_a = |e|^{-1}\epsilon^{ij}\tau_{aj}$.

\subsection{Anomalies}
\label{anomalies}

Let us now consider the Ward identities satisfied by the stress
tensor \eqref{stress-tens}. First of all, its trace is given by
\begin{equation}
T = {e^a}_i {T^i}_a = \pi\ell\,[2a+\alpha_3 A]\,R - 2\pi\alpha_3\nabla_i\omega^i\,,
                      \label{traceT}
\end{equation}
where $R$ denotes the scalar curvature of the boundary. To obtain \eqref{traceT},
we used $k_{[ij]}=0$ and equ.~\eqref{curv}, which implies
\begin{displaymath}
R = \frac 4{\ell^2|e|}\epsilon^{ij}\epsilon_{ab}\,{e^a}_i {k^b}_j\,.
\end{displaymath}
Using the central charges
\begin{equation}
c_L = 24\pi\left[a\ell + \alpha_3\left(\frac{A\ell}2-1\right)\right]\,, \qquad
c_R = 24\pi\left[a\ell + \alpha_3\left(\frac{A\ell}2+1\right)\right]\,, \label{central}
\end{equation}
of the dual conformal field theory, obtained in \cite{Blagojevic:2004hj} by
computing the Poisson bracket algebra of the asymptotic symmetry generators,
\eqref{traceT} can be rewritten as
\begin{equation}
T = \frac{c_L+c_R}{24}R - 2\pi\alpha_3\nabla_i\omega^i\,. \label{traceTc}
\end{equation}
The first piece is the usual covariant expression for the trace anomaly,
whereas the second one transforms non-covariantly under local Lorentz
transformations. We will come back to this point later.\\
The energy-momentum tensor \eqref{stress-tens} is not symmetric,
\begin{equation}
T_{ab}-T_{ba}=2\pi\alpha_3\,^*R_{ab} = \frac{c_R-c_L}{24}\,^*R_{ab}\,,
\label{Lorentz-anomaly}
\end{equation}
where $T_{ab}= e_{ai}{T^i}_b$, and $^*R^{ab}= (2|e|)^{-1}\epsilon^{ij}{R^{ab}}_{ij}$ is
the Hodge dual of the Riemann tensor. \eqref{Lorentz-anomaly} means that there is
a Lorentz anomaly in the dual field
theory \cite{Alvarez-Gaume:1983ig,Bardeen:1984pm,Alvarez-Gaume:1984dr,Ginsparg:1985qn}:
Under an infinitesimal local Lorentz transformation the zweibein transforms as
$\delta_{\alpha}{e^a}_i = -{\alpha^a}_b {e^b}_i$, so the variation of the quantum
effective action is $\delta_{\alpha}\Gamma_{\mathrm{eff}}=-\int d^2x{\alpha^a}_b
{e^b}_i(\delta\Gamma_{\mathrm{eff}}/\delta {e^a}_i)$. But $e_{bi}
(\delta\Gamma_{\mathrm{eff}}/\delta {e^a}_i)=|e|\,T_{ba}/2\pi$,
so one has
\begin{displaymath}
\delta_{\alpha}\Gamma_{\mathrm{eff}} = -\frac 1{2\pi}\int d^2x |e|\alpha^{ab}T_{ba}\,.
\end{displaymath}
Since $\alpha^{ab}$ is antisymmetric, it follows that the non-invariance of the
effective action under local Lorentz transformations is equivalent to asymmetry of
$T_{ab}$.\\
Let us finally compute the divergence of \eqref{stress-tens}. Making use of
\eqref{dk}, one obtains
\begin{equation}
\nabla_i{T^i}_a = \pi\alpha_3 R {e_a}^j\omega_j\,, \label{divT}
\end{equation}
where $\nabla_i$ denotes the covariant derivative with respect to both local Lorentz
transformations and diffeomorphisms, i.~e.
\begin{displaymath}
\nabla_i{T^i}_a = \partial_i{T^i}_a + {\Gamma^i}_{ij}{T^j}_a - \omega^b_{i\,a}{T^i}_b\,.
\end{displaymath}
To see that \eqref{divT} is the correct Ward identity,
observe that under an infinitesimal coordinate transformation
$x^i\mapsto x^i - \xi^i$, the zweibein varies as
$\delta_{\xi}{e^a}_i = {e^a}_j{\tilde\nabla}_i\xi^j + \xi^j{\tilde\nabla}_j{e^a}_i$,
with ${\tilde\nabla}_j{e^a}_i = \partial_j{e^a}_i - {\Gamma^k}_{ji}{e^a}_k$ being the
covariant derivative w.~r.~t.~diffeomorphisms. Using $\delta\Gamma_{\mathrm{eff}}/
\delta {e^a}_i = |e|\,{T^i}_a/2\pi$, the variation of the effective action becomes
\begin{displaymath}
\delta_{\xi}\Gamma_{\mathrm{eff}} = \frac1{2\pi}\int d^2x |e|\,{T^i}_a({e^a}_j
{\tilde\nabla}_i\xi^j + \xi^j{\tilde\nabla}_j{e^a}_i)\,.
\end{displaymath}
Integrating by parts the first term, using ${T^i}_j = {T^i}_a{e^a}_j$ and
${\tilde\nabla}_j{e^a}_i = -\omega^a_{j\,b}{e^b}_i$ (which follows from
$\nabla_j{e^a}_i=0$), one finally gets
\begin{equation}
\delta_{\xi}\Gamma_{\mathrm{eff}} = \frac1{2\pi}\int d^2x |e|\,\xi^j\left[-\nabla_i
{T^i}_j + \omega^{ab}_j\,T_{ab}\right]\,.
\end{equation}
Invariance under diffeomorphisms implies then
\begin{equation}
\nabla_i{T^i}_j = \omega^{ab}_j\,T_{ab}\,. \label{divT-gen}
\end{equation}
If Lorentz symmetry is preserved so that $T_{ab}$ is symmetric, the term on the
r.~h.~s.~vanishes due to the antisymmetry of the spin connection, and one has
the usual conservation law $\nabla_i{T^i}_j = 0$. In our case, however, Lorentz
symmetry is broken, and the antisymmetric part of $T_{ab}$ is given by
\eqref{Lorentz-anomaly}. Plugging this into \eqref{divT-gen} yields exactly
\eqref{divT}. This means that in the field theory dual of \eqref{MBaction},
diffeomorphism invariance is preserved. Note that, by adding local counterterms,
it is always possible to shift the Lorentz anomaly into a diffeomorphism anomaly
and vice-versa \cite{Bardeen:1984pm}.\\
As we said earlier, the trace \eqref{traceTc} of the stress tensor is not covariant.
This is a general feature of anomalies: There are consistent and covariant
anomalies \cite{Bardeen:1984pm}. The former satisfy the Wess-Zumino consistency
conditions \cite{Wess:1971yu} and the corresponding currents are obtained by varying
the vacuum functional with respect to the gauge potential, whereas the latter
are obtained by adding to the corresponding consistent anomaly a local function
of the gauge potential (the so-called Bardeen-Zumino polynomial). The resulting
current is covariant under local gauge transformations.
In our case, by adding to the energy-momentum tensor \eqref{stress-tens} the
Bardeen-Zumino polynomial
\begin{equation}
{{\cal P}^i}_a = -\frac{2\pi\alpha_3}{|e|}\epsilon^{ij} e_{am}\nabla_j(^*\omega)^m\,,
\end{equation}
we get the covariantly transforming stress tensor ${\tilde T}^i_{\;\;a} = {T^i}_a
+ {{\cal P}^i}_a$, whose trace and divergence are given respectively by
\begin{equation}
\tilde T = \frac{c_L+c_R}{24}R\,, \qquad \nabla_i\,{\tilde T}^i_{\;\;a} = 0\,.
\end{equation}
For the antisymmetric part of ${\tilde T}_{ab}$ one gets
\begin{equation}
{\tilde T}_{ab} - {\tilde T}_{ba} = 4\pi\alpha_3\,^*R_{ab}\,,
\end{equation}
which is twice the right hand side of \eqref{Lorentz-anomaly}. Observe that
${\tilde T}^i_{\;\;a}$ is exactly the result we would have obtained by
dropping the contribution of the last piece in \eqref{deltaItot}, i.~e.~, by
considering the zweibein and the spin connection as independent fields.

\subsection{Chern-Simons gauge transformations}

A particular example resolving the constraints \eqref{dk}, \eqref{tors=0} and
\eqref{curv} is given by
\begin{eqnarray}
e^0 &=& \frac{\ell}2(du-dv)\,, \qquad e^1 = \frac{\ell}2(du+dv)\,, \qquad \omega = 0\,,
\nonumber \\
k^0 &=& 2G[-{\tilde L}(u)du + L(v)dv]\,, \qquad k^1 = 2G[{\tilde L}(u)du + L(v)dv]\,,
 \label{ban-sol}
\end{eqnarray}
where $u=(x+t)/\ell$, $v=(x-t)/\ell$ are light-cone coordinates on the boundary,
$L(v)$ and ${\tilde L}(u)$ denote arbitrary functions, and $G$ is the
3d Newton constant. The corresponding
three-dimensional line element reads
\begin{equation}
d{\hat s}^2 = 4G\ell\,({\tilde L}du^2+Ldv^2) + (\ell^2 e^{2\rho} + 16 G^2 L{\tilde L}
e^{-2\rho})\,dudv + \ell^2 d\rho^2\,.
\end{equation}
\eqref{ban-sol} represents a generalization to nonvanishing torsion of the general
solution with flat boundary geometry obtained in \cite{Banados:1998gg}. $L$ and
$\tilde L$ describe right- and left-moving gravitational waves on AdS$_3$ respectively.
Using \eqref{ban-sol} in \eqref{stress-tens} yields the holographic stress tensor
\begin{equation}
T_{vv} = \frac{2Gc_R}{3\ell}L(v)\,, \qquad T_{uu} = \frac{2Gc_L}{3\ell}{\tilde L}(u)\,,
\qquad T_{uv} = T_{vu} = 0\,. \label{stress-tens-ban}
\end{equation}
In the case $\alpha_3=\alpha_4=0$, when $c_L=c_R=3\ell/2G$ \cite{Brown:1986nw},
this reduces to $T_{vv}=L$, $T_{uu}=\tilde L$, as it must be \cite{Banados:1998gg}.\\
The Chern-Simons connections corresponding to the solution \eqref{ban-sol} are
\begin{eqnarray}
A^0_v &=& -e^{\rho} + e^{-\rho}\frac{4GL}{\ell}\,, \qquad
A^1_v = e^{\rho} + e^{-\rho}\frac{4GL}{\ell}\,, \qquad
A^2_{\rho} = 1\,, \label{conn-L} \\
{\tilde A}^0_u &=& -e^{\rho} + e^{-\rho}\frac{4G{\tilde L}}{\ell}\,, \qquad
{\tilde A}^1_u = -e^{\rho} - e^{-\rho}\frac{4G{\tilde L}}{\ell}\,, \qquad
{\tilde A}^2_{\rho} = -1\,, \nonumber
\end{eqnarray}
and all other components vanishing. We may now ask which gauge transformations
preserve this form of the connection. Under an infinitesimal gauge transformation
the connection $A$ changes according to 
\begin{displaymath}
\delta A = -du - [A,u]\,,
\end{displaymath}
where $u=u^A\tau_A$ is an sl$(2,\bR)$-valued scalar. One finds that the form
\eqref{conn-L} is preserved iff
\begin{eqnarray}
u^0 &=& -\alpha(v)\,e^{\rho} + \left[\frac{4GL}{\ell}\alpha(v)-\frac{\alpha''(v)}2
        \right]e^{-\rho}\,, \nonumber \\
u^1 &=& \alpha(v)\,e^{\rho} + \left[\frac{4GL}{\ell}\alpha(v)-\frac{\alpha''(v)}2
        \right]e^{-\rho}\,, \nonumber \\
u^2 &=& -\alpha'(v)\,, \nonumber
\end{eqnarray}
where $\alpha(v)$ denotes an arbitrary function. The variation of $L$ is
\begin{displaymath}
\delta L = -2\alpha'(v)L - \alpha(v)L' + \frac{\ell}{8G}\alpha'''(v)\,,
\end{displaymath}
which implies
\begin{equation}
\delta T_{vv} = -2\alpha'(v)T_{vv} - \alpha(v)T_{vv}' + \frac{c_R}{12}\alpha'''(v)
                \label{transf-Tvv}
\end{equation}
for the component $T_{vv}$ of the stress tensor. \eqref{transf-Tvv} is the correct
transformation law under conformal transformations, and confirms that $c_R$ is the
central charge of the right-moving sector. An analogous calculation for $\tilde A$
yields the transformation law for $T_{uu}$ with anomaly proportional to $c_L$.
Note that one has $c_R=6t$, $c_L=6\tilde t$, where $t$ and
$\tilde t$ denote the Chern-Simons coupling constants \eqref{t}.

\section{Entropy of the Riemann-Cartan black hole}
\label{entr-RC}

If we choose
\begin{displaymath}
L(v) = \frac{m\ell-j}2\,, \qquad {\tilde L}(u) = \frac{m\ell+j}2\,,
\end{displaymath}
where $m$ and $j$ are constants, and change the coordinates according to
\begin{displaymath}
e^{2\rho} = \frac 12\left[\sqrt{\frac{r^4}{\ell^4} - 8Gm\frac{r^2}{\ell^2} +
\frac{16G^2j^2}{\ell^2}} + \frac{r^2}{\ell^2} - 4Gm\right]\,, \qquad
u = \phi + \frac t{\ell}\,, \qquad v = \phi - \frac t{\ell}\,,
\end{displaymath}
\eqref{ban-sol} reduces to the so-called Riemann-Cartan (RC) black
hole \cite{Garcia:2003nm}, whose metric is identical to that of the BTZ solution,
\begin{equation}
d{\hat s}^2 = -N^2 dt^2 + \frac{dr^2}{N^2} + r^2(d\phi + N^{\phi}dt)^2\,,
\end{equation}
with
\begin{displaymath}
N^2 = -8Gm + \frac{r^2}{\ell^2} + \frac{16G^2j^2}{r^2}\,, \qquad N^{\phi} =
\frac{4Gj}{r^2}\,.
\end{displaymath}
Note that the spin connection is different from the Christoffel connection due to
nonvanishing torsion.\\
The holographic stress tensor \eqref{stress-tens-ban} corresponding to the RC
black hole is given by
\begin{equation}
T_{vv} = \frac{Gc_R}{3\ell}(m\ell-j) \equiv T_0\,, \qquad
T_{uu} = \frac{Gc_L}{3\ell}(m\ell+j) \equiv {\tilde T}_0\,.
\end{equation}
$T_0$ and ${\tilde T}_0$ are the zero-modes in a Fourier expansion of the
energy-momentum tensor. The mass and angular momentum of the solution are
\begin{eqnarray}
M &=& \frac 1{\ell}(T_0 + {\tilde T}_0) = m + \frac{\alpha_3}a\left(\frac{Am}2 -
\frac j{\ell^2}\right)\,, \nonumber \\
J &=& {\tilde T}_0 - T_0 = j + \frac{\alpha_3}a\left(\frac{Aj}2 - m\right)\,. \label{MJ}
\end{eqnarray}
The conserved charges \eqref{MJ} coincide with the ones computed
in \cite{Garcia:2003nm,Blagojevic:2005pd}. For AdS$_3$ in global coordinates,
which represents the ground state and corresponds to $j=0$, $8Gm=-1$, one gets
\begin{displaymath}
M_{{\mathrm{AdS}}_3}\ell = -2\pi\ell\left[a + \alpha_3\frac A2\right] =
-\frac{c_R+c_L}{24}\,, \qquad J_{{\mathrm{AdS}}_3} = 2\pi\alpha_3 = \frac{c_R-c_L}{24}\,.
\end{displaymath}
The nonvanishing ground state angular momentum is due to the asymmetry of the
central charges, which prevents the left- and right-moving zero point momenta
from cancelling each other \cite{Kraus:2005zm}.\\
The entropy of the RC black hole
was obtained in \cite{Blagojevic:2006jk} by calculating the Euclidean action,
with the result
\begin{equation}
S = \frac{2\pi r_+}{4G} + 4\pi^2\alpha_3\left(Ar_+ - 2\frac{r_-}{\ell}\right)\,,
\label{entropy}
\end{equation}
where 
\begin{displaymath}
r_{\pm}^2 = 4Gm\ell^2\left[1\pm\sqrt{1 - \frac{j^2}{m^2\ell^2}}\right]
\end{displaymath}
are the locations of the outer and inner horizon. The first term in \eqref{entropy}
is the standard Bekenstein-Hawking result, proportional to the area of the event
horizon, whereas the second term represents a correction due to the other terms
in the action \eqref{MBaction}. The quantities $S, M, J$ satisfy the first law
of thermodynamics \cite{Blagojevic:2006jk}
\begin{displaymath}
dM = TdS + \Omega dJ\,,
\end{displaymath}
with the Hawking temperature $T$ and the angular velocity of the horizon $\Omega$
given by
\begin{displaymath}
T = \frac{r_+^2 - r_-^2}{2\pi\ell^2r_+}\,, \qquad \Omega = \frac{4Gj}{r_+^2}\,.
\end{displaymath}
Using the central charges \eqref{central} and the conformal weights $T_0$,
${\tilde T}_0$ in the Cardy formula yields the microscopic entropy
\begin{equation}
S_{\mathrm{micr}} = 2\pi\sqrt{\frac{c_R T_0}6} + 2\pi\sqrt{\frac{c_L{\tilde T}_0}6}\,,
\label{cardy}
\end{equation}
which agrees exactly with the thermodynamic entropy \eqref{entropy}. This was first
shown in \cite{Blagojevic:2006hh}. Note that the derivation of the Cardy formula
uses modular invariance of the CFT partition function (see e.~g.~\cite{Carlip:1998qw}),
which requires $c_R-c_L$ to be a multiple of 24\cite{Polchinski:1998rq}, i.~e.~, one
must have
\begin{equation}
2\pi\alpha_3 \in \bZ. \label{quant-cond}
\end{equation}
Note in this context that in Euclidean signature, the gauge group in the CS formulation
of the MB model becomes SL$(2,\bC)$, with maximal compact subgroup SU$(2)$, so that
$\alpha_3$ is subject to a topological quantization
condition \cite{Witten:1989ip,Cacciatori:2005wz}, which might be related
to \eqref{quant-cond}.

\acknowledgments

We are grateful to Sergio Cacciatori, St\'ephane Detournay and Rodrigo Olea for
useful discussions, and to Finn Larsen for clarifying correspondence.
This work was partially supported by INFN, MURST and by the
European Commission program MRTN-CT-2004-005104.

\normalsize

\appendix

\section{Wess-Zumino consistency conditions}

In this appendix we shew that the anomalies \eqref{traceTc} and \eqref{Lorentz-anomaly}
satisfy the Wess-Zumino consistency conditions \cite{Wess:1971yu}. It was shown in
section \ref{anomalies} that under an infinitesimal local Lorentz transformation
$\alpha^{ab}$, the vacuum functional changes as
\begin{equation}
\delta_{\alpha}\Gamma_{\mathrm{eff}} = \frac 1{2\pi}\int d^2x |e|\alpha^{ab}T_{ab}\,.
\label{del-alph}
\end{equation}
Let us assume that the Lorentz anomaly takes the form $\epsilon^{ab}T_{ab}=\beta R$
for some constant $\beta$ (in our case $\beta=-\pi\alpha_3$). Under an infinitesimal
local Weyl transformation $\delta_{\varphi}{e^a}_i=\varphi{e^a}_i$,
\eqref{del-alph} varies as
\begin{equation}
\delta_{\varphi}\delta_{\alpha}\Gamma_{\mathrm{eff}} = -\frac{\beta}{\pi}\int d^2x |e|
\alpha\nabla^2\varphi\,, \label{del-phi-alph}
\end{equation}
where the function $\alpha$ is defined by $\alpha^{ab}=\alpha\epsilon^{ab}$.
On the other hand, applying first a Weyl transformation yields
\begin{equation}
\delta_{\varphi}\Gamma_{\mathrm{eff}} = \frac 1{2\pi}\int d^2x |e|\,T\varphi\,.
\label{del-phi}
\end{equation}
Under the assumption $T=\gamma R+\tilde\gamma\nabla_i\omega^i$, with $\gamma,\tilde\gamma$
constants (in our case $\gamma=(c_L+c_R)/24$, $\tilde\gamma=-2\pi\alpha_3$),
\eqref{del-phi} splits into two pieces, the first of which being Lorentz invariant,
whereas the second gives the variation
\begin{equation}
\delta_{\alpha}\delta_{\varphi}\Gamma_{\mathrm{eff}} = -\frac{\tilde\gamma}{2\pi}\int
d^2x |e|\,\varphi\nabla^2\alpha\,, \label{del-alph-phi}
\end{equation}
and we used $\delta\omega=-d\alpha$. As Weyl- and Lorentz transformations commute,
\eqref{del-alph-phi} and \eqref{del-phi-alph} must be the same. Integrating by parts
twice yields then the relation
\begin{equation}
\tilde\gamma=2\beta\,,
\end{equation}
which is indeed satisfied by the anomalies \eqref{traceTc}, \eqref{Lorentz-anomaly}.

\end{document}